\documentstyle[psfig,numreferences,editedvolume]{crckapb} 

\def\la{Ly$\alpha$}
\def\hb{H$\beta$}

\def\aa{{\rm A$\,$\&$\,$A}}            
\def\apj{{\rm ApJ}}                    
\def\aj{{\rm AJ}}                      
\def\mnras{{\rm MNRAS}}                        

\def\Msun{\thinspace\hbox{$\hbox{M}_{\odot}$}}

\def\kpc{\thinspace\hbox{kpc}}

\begin{opening}
\title{Gas-rich local dwarf star-forming galaxies \protect \\
	and their connection with the distant
Universe} \protect \\

\author{D. Kunth}
\institute{Institut d'Astrophysique de Paris\\
           98 bis Boulevard Arago \\
	   75014 Paris, France}

\end{opening}

\runningtitle{Gas-rich dwarf star-forming galaxies}

\begin{document}


\begin{abstract}

I discuss the properties of gas-rich forming galaxies. I particularly
emphasize the latest results on \la\ emission that are relevant to the 
search of distant young galaxies. The interdependance of the \la\ escape 
with the
properties of the ISM in starburst galaxies is outlined. A new model from
G. Tenorio-Tagle and his collaborators explain \la\ profiles in starburst
 galaxies from the
hydrodynamics of superbubbles powered by massive stars. I  stress again 
that since  \la\
is primarely a diagnostic of the ISM, it is mandatory to
understand how the ISM and \la\ are related to firmly relate
\la\ to the cosmic star--formation rate. 
\end{abstract}

\section{Introduction}

Among galaxies that pertain to our local Universe the gas-rich ones and in
particular the ones with violent episodes of massive star formation attract
our attention. Along the Hubble sequence the fractional gas content ($i.e.$
the gas to total mass ratio) increases as one moves from early types
to late types and reaches values as large as 0.1 to 0.5 for Irr galaxies. 
More extreme cases are found amongst LSBs (Low surface brightness
galaxies) and the dwarfs star-forming galaxies in which the total amount
of gas (mainly atomic gas as the molecular gas content in these low mass
objects is still unknown) could be the dominant component.
Our recent view of the distant Universe opens the possibility that small
gaseous systems gradually merge into large ones. Galaxy formation appears
more gradual that had been anticipated and one would expect to see some left 
over debris in our local Universe after a priviledged epoch of star
formation at around z of 1-2 . Dwarf
star-forming galaxies naturally comes into discussion
 for at least two reasons -
both linked to the distant Universe. First they appear so chemically
unevolved that they represent the best site to look at for the determination
of the primordial
Helium abundance. Indeed, their measured helium abundances
  are accurately
 determined and
are close to the primordial helium value since the
He products from stellar burning is presumably minimal. 
Second they offer ideal local counterparts of what should be distant
protogalaxies at the onset of their first massive star formation episodes.
I will focus on this later point in the rest of this review.

\section{The  dwarf star-forming galaxies}

\subsection{Their overall properties}

Dwarf star-forming galaxies are sometimes called Blue Compact Dwarf
(BCD) or HII galaxies 
because their ongoing star-formation rate  large with respect to
their integrated past star formation rate so that their
overall spectroscopic appearance is that of a Giant HII region. Although
HII galaxies also comprise a wider class of objects in term of
mass and luminosity I shall refer to this terminology in the following
 discussion.
Local HII galaxies have absolute luminosities dominated by the
young stellar population and  range from M$_{B}$=-13
to -17 (if only restricted to dwarfs). They are bluer than ordinary irregular
galaxies  (with
typically U-B=-0.6 and B-V=0.0) and are rich in ionized gas. The
equivalent width of the \hb\ line ranges from 30 to 200 \AA\ implying 
a burst star formation  no older than a few Myrs. The role of massive stars
is to energize the ISM and enrich it with heavy metals and some helium.
These galaxies are chemically unevolved and
 their metal distribution  peaks at 1/10th the solar value, with the
most deficient galaxy  IZw18 at only 1/50th. Their neutral gas content is
relatively high, amounting a few 10$^{7}$ \Msun\ in general, $i.e.$ a large
fraction  of the luminous mass. On the other hand there are some indications
that the dynamical mass is much larger, \cite{CB}. UV to 
IR studies  clearly indicate that massive star formation is a discontinuous
process concentrated to short bursts intervened by longer passive periods.
 Their UV spectra
 show small amount of dust  that
makes them ideal targets to study locally.

\subsection{Their link with the primeval galaxies}

The major motivation in
follow-up studies of  unevolved and nearly dust-free
 HII galaxies is  their possible ressemblance
with distant protogalaxies in their early phase of
star-formation. Assuming that protogalaxies would undergo a
rapid collapse on a short dynamical timescale, \cite{E}, the
expectation is that an L$^{*}$ galaxy could be detected from its redshifted
\la\ emission line. Indeed in a dust free case and a normal IMF such a
 galaxy could
pop up with a star formation rate of nearly 1000 \Msun/yr producing a
\la\ luminosity of 10$^{45}$ erg/sec (this
 amounts
few percent of the total luminosity). At a redshift of 3 this translates to
10$^{-15}$ erg/cm$^{2}$/sec, in principle easily detectable even from 
deep spectrocopic
searches or narrow band imaging using panoramic CCDs. The failure to
detect such a strong  \la\ is an additional piece of evidence that the
overall process of galaxy formation is more complex: galaxies form from
smaller building blocks and/or 
luminous protogalaxies have to be found at much larger redshift and/or are
enshrouded in large dusty cocoons.

\section{The Early IUE observations and interpretations}
While galaxies at their very early stage could be nearly dust-free hence
easily detectable from their \la\ emission \cite{PP}; early ultraviolet 
observations of nearby HII galaxies, 
have 
revealed  a  much weaker \la\ emission than
 predicted by starburst models and simple models 
of galaxy formation. In some other galaxies \la\ was non-existent or even 
appeared as a broad absorption profile
(\cite{DJK,H84,H88,K97,K98,MT,TDT}). 
 For young star-forming galaxies without so much dust
as to  suppress \la\ , large equivalent widths are expected in the
 range of
100-200 (1+z)~\AA \ , \cite{CF}. 
However it was early realized that pure extinction by dust would be unable to
explain the low observed \la /\hb\  (although Calzetti and Kinney 
 \cite{CKI} tentatively proposed 
that proper extinction laws would
correctly match the predicted recombination value). 
Valls-Gabaud \cite{VG} on
the other hand suggested that ageing  starbursts could reduce 
\la\ equivalent widths because they are affected
by strong underlying  stellar atmospheric absorptions.  Early IUE data have 
provided evidence for an anticorrelation between the  \la /\hb\ ratio and
 the HII galaxy metallicity. These results were
attributed to the effect of resonant scattering of \la\ photons
and their subsequently increased absorption by dust (\cite{CF,CN,GK,MT} 
and references therein).

Finally  \cite{CF} advocated that
the structure of the interstellar medium (porosity and multi-phase structure) 
is most probably an important factor for the visibility of the \la \ line.

\section{HST observations}

\begin{figure} 
\psfig{file=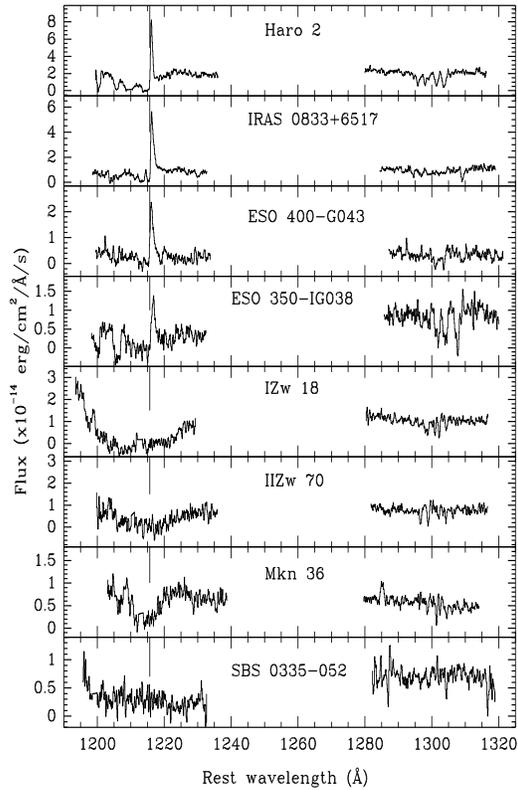,height=12cm,width=8cm}
\caption[]{
GHRS spectra of all the galaxies in \cite{K98}. The spectra have been
shifted to rest velocity assuming the redshift derived from optical
emission lines. Vertical bars indicate the wavelength at which the Ly$\alpha$\
emission line should be located. The geocoronal emission profile has been
truncated for the sake of clarity. The spectra have been plotted after
rebinning to 0.1~\AA\ per pixel and smoothed by a 3 pixel box filter.  }
\label{fig:total}
\end{figure}

New  observations performed with the HST
indicate that the velocity structure in the
interstellar medium plays a key role in the transfer and escape
 of Ly$\alpha$\
photons. Kunth et al. \cite{K94}
and Lequeux et al. \cite{LKM} have used the Goddard High Resolution
 Spectrograph
(GHRS) onboard the Hubble Space Telescope (HST) to observe
  Ly$\alpha$\ and the interstellar lines 
 O\,{\sc I} 1302.2~\AA
\ and Si\,{\sc II} 1304.4~\AA\ .   The O\,{\sc I} 1302~\AA\ and 
Si\,{\sc II} 1304~\AA\  allow to
measure with reasonable accuracy the mean velocity at which the absorbing
material lies with respect to the star-forming region of a given galaxy.
Surprisingly, Ly$\alpha$\ was observed only in absorption in  IZw~18, the
 most metal--poor starburst galaxy known, \cite{K94}. 
 Meanwhile to add to the confusion, a  Ly$\alpha$\ emission line showing
 a clear P--Cygni component, has been detected in  the star-forming galaxy 
Haro~2,  
 dustier than IZw~18 and with  $Z=1/3 ~Z_{\odot}$ , \cite{LKM}.
The detection of such a profile in the
 Ly$\alpha$\
emission line led us to postulate that the line was visible
because the absorbing neutral gas was velocity--shifted with respect to the
ionized gas. This was confirmed by the analysis of the UV O\,{\sc I} and 
Si\,{\sc II}
absorption lines (blue--shifted by 200~km\thinspace
 s$^{-1}$\  with respect
 to the optical emission lines) and that of the profile of the H$\alpha$ line
\cite{LKML}.
Observations were subsequently made on additional  galaxies  by Kunth et al.
\cite{K98} while Thuan and Isotov \cite{TI} have obtained GHRS 
spectra of two more starburst galaxies,
namely Tol65 and T1214-277. Tol65 reveals a broad damped \la\ absorption
 while T1214-277 
shows a pure \la\ emission profile with an equivalent width of 70 \AA \ 
and with no blue absorption. 
The individual spectra of galaxies observed in \cite{K98} are shown in
Fig.~\ref{fig:total}.


\section{The role of the velocity structure of the ISM}

 Three types of observed lines have been identified so far: pure \la\ 
emission;  broad damped  \la\ absorption  centered at the 
wavelength corresponding to the redshift of the HII emitting gas
  and \la\
emission with blue shifted absorption features, leading in some cases 
to P--Cygni profiles.

\begin{figure} 
\psfig{file=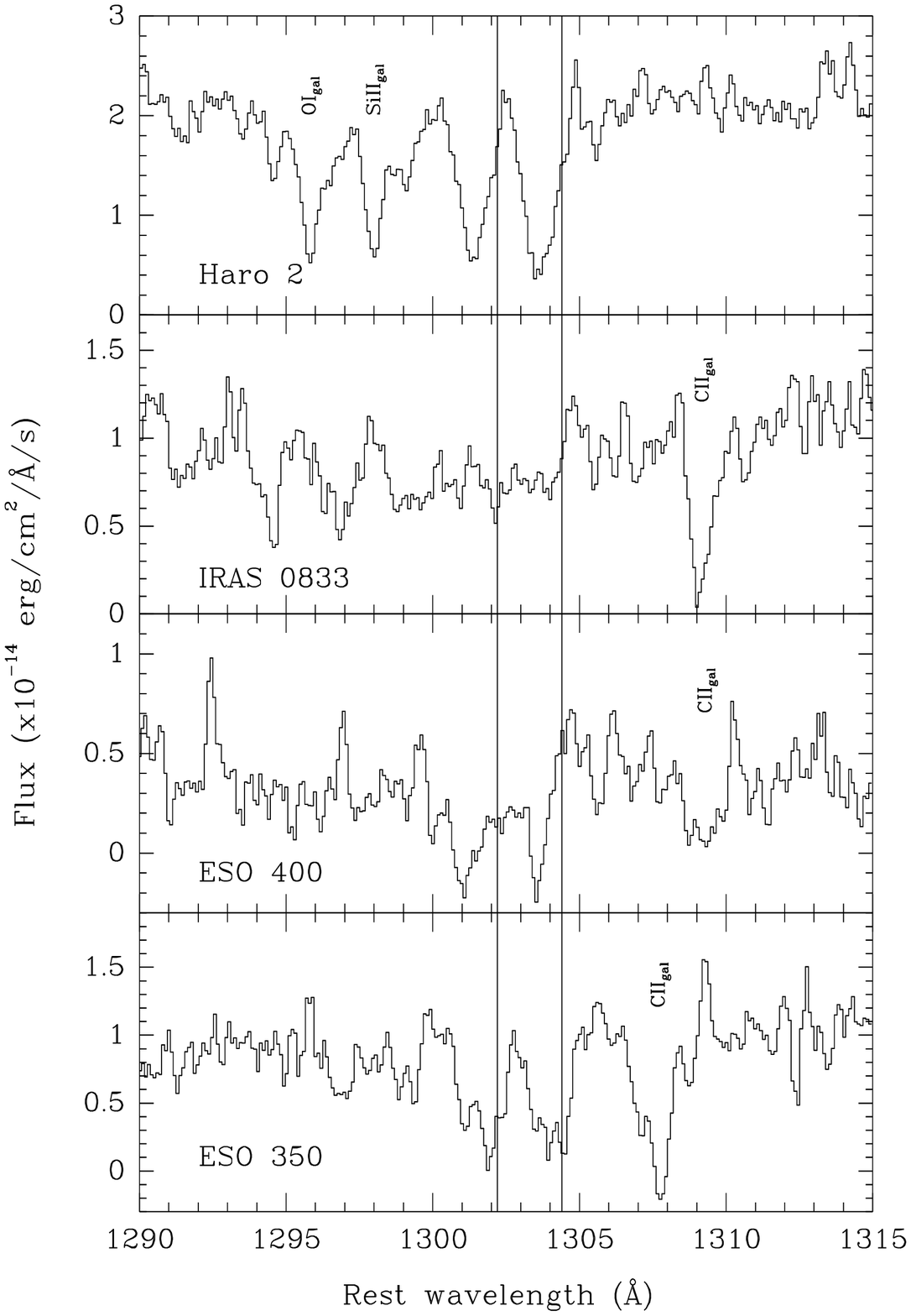,height=10cm,width=12cm}
\caption[]{
Detail of the O\,{\sc I} and Si\,{\sc II} region for the galaxies
 with Ly$\alpha$\
emission. The vertical bars indicate the wavelength at which the O\,{\sc I} and
Si\,{\sc II} absorption lines should be located, according to the redshift
 derived
from optical emission lines. Some Galactic absorption lines have been
marked. Note that the metallic lines appear systematically blueshifted in
these galaxies with respect to the systemic velocity. In some cases there
is no significant absorption at all at zero velocity. }
\label{fig:oila}
\end{figure}

\begin{figure} 
\psfig{file=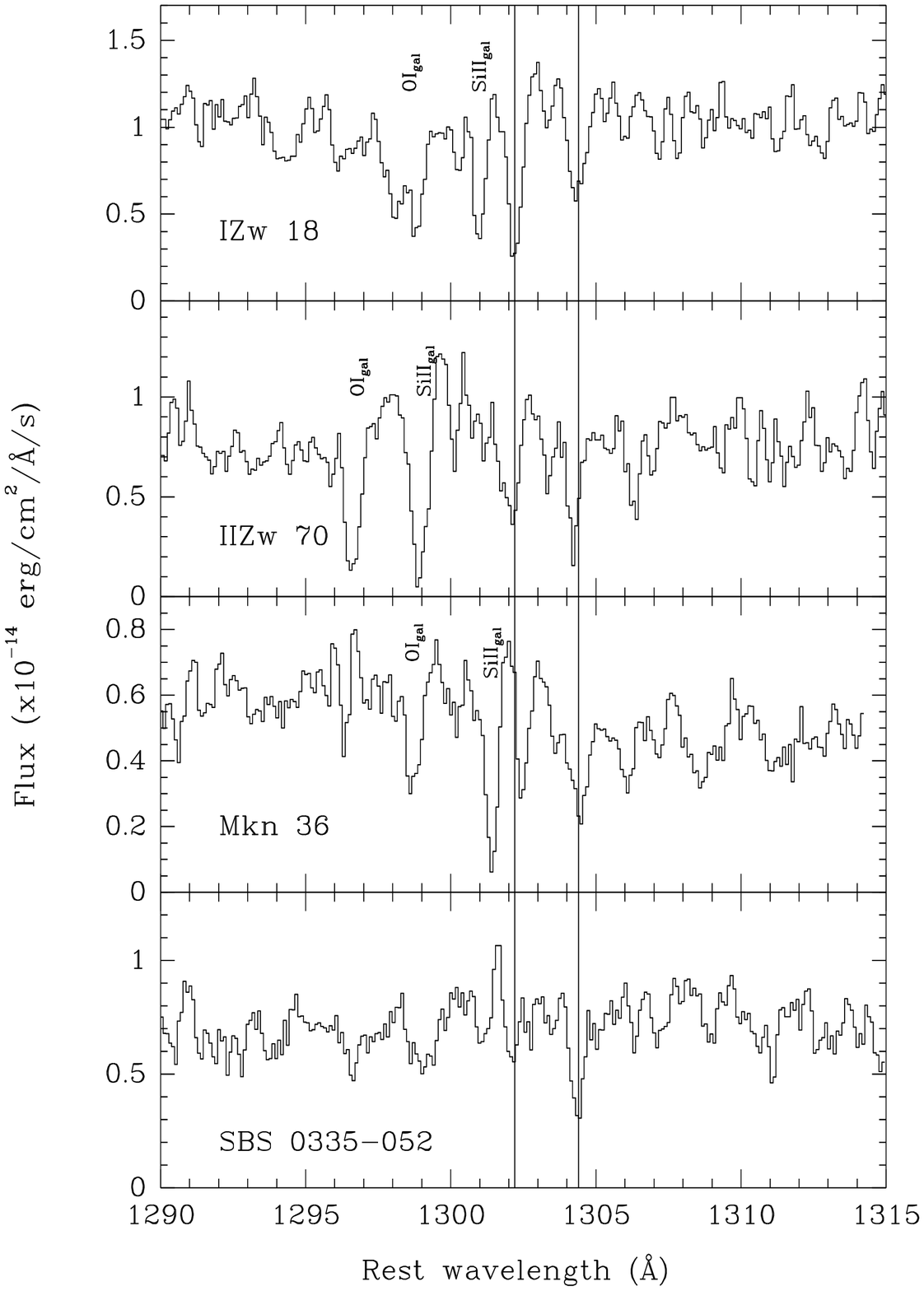,height=10cm,width=12cm}
\caption[]{
O\,{\sc I} and Si\,{\sc II} region for the galaxies showing damped Ly$\alpha$\
absorptions. Details as in Fig.~\ref{fig:oila}. Note that in these
galaxies the metallic lines are essentially at the same redshift as the
ionized gas, indicating the presence of static clouds of neutral gas, as
discussed in the text. }
\label{fig:oiab}
\end{figure}

 \la\ emission with deep blueward absorption troughs
 evidence a wide velocity field. The equivalent widths
 of the \la\ emission range
between 10 and 37~\AA \ hence much below the value predicted
by Charlot and Fall, \cite{CF} for a dust--free starburst model. 
In all cases, interstellar absorption lines (OI, SiII) are
 significantly blueshifted with respect to the HII gas
 (see Fig.~\ref{fig:oila}).
On the other hand, if the HI is static with 
respect to HII, the destroyed \la\ 
photons are those emitted by the HII region and the interstellar
lines are not displaced (see Fig.~\ref{fig:oiab}). 

 In galaxies with low dust
content - IZw~18 has a dust-to-gas ratio  at
 least 50 times smaller than the Galactic
value, \cite{K94} -, it remains possible to weaken
observationally  \la\ by simple multiple resonant scattering from a static
neutral gas and  produce an absorption feature. The H\,{\sc I} cloud 
surrounding these galaxies might be leaking  \la\ 
photons through its external surface. The  \la\  line would then become
 very
hard to detect because of its low surface brightness.
On the other hand, the situation in Haro~2 is quite different.
The \la\ P~Cygni profile suggests a rather modest amount of neutral gas 
 of the order of
N~(H\,{\sc I}) = 7.7x$10^{19}$ atoms cm$^{-2}$. The crucial point here
 is that the
neutral gas responsible for the absorption  is not at the
velocity at which the \la\ photons were emitted but is being pushed out by an
 expanding
envelope around the H\,{\sc II} region,  at velocities close to 200
km\thinspace s$^{-1}$. This interpretation is  strengthened by
 the presence of the
 blueshifted absorptions of O\,{\sc I}, Si\,{\sc II} and Si\,{\sc III}
  precisely due to 
outflowing gas in
front of the ionizing hot stars, 
(see also Legrand et al., \cite{LKML}). 
Data on other  H\,{\sc II} galaxies with detected \la\ emission
confirm that Haro 2 is not an isolated case. Most spectra show
\la\ emission with a broad absorption on their blue side except
 for ESO 350-IG038 in
which the emission is seen atop of a broad structure requiring several
filaments.
 When  metallic lines are detected, they are always
blueshifted with respect to the ionized gas, further supporting the
interpretation. In the case of ESO 350-IG038 the velocity structure seems
to be more complicated and several components at different velocities are
identified on the metallic lines. Note that there might be a secondary peak 
emission in the
 blue side of the main line in the spectrum of ESO-400-G043.


The main conclusion drawn from this set of data is that complex
velocity structures are determining the Ly$\alpha$\ emission line 
detectability,
showing the strong energetic impact of the star-forming regions onto their
surrounding ISM. 
 We want to stress
that this effect seems to be almost independent of
the dust and metal abundance of the gas. 

 Thuan and Isotov \cite{TI} have 
detected strong Ly$\alpha$\ emission in T1214-277, with no clear evidence of
blueshifted Ly$\alpha$\ absorption. They argue that a significant fraction
 of the area covered by the slit along
the line of sight is essentially free from neutral gas, suggesting a patchy
or filamentary structure of the neutral clouds. Such a geometry would be
unlikely  in galaxies surrounded by enormous H\,{\sc I} clouds, as 
in IZw~18 and similar objects. 

\section{The evolution of superbubbles in extended HI halos}

\begin{figure} 
\psfig{file=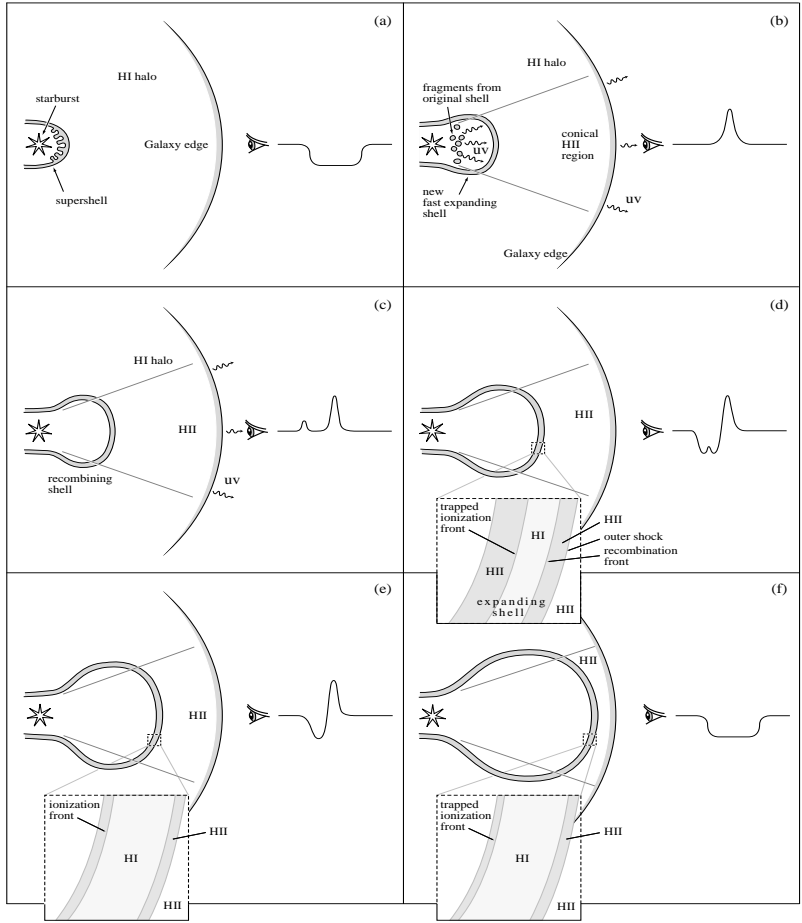,height=15cm,width=12cm}
\caption[]{The evolution of a superbubble as a function of time.}
\label{fig:superbubble}
\end{figure}

To account naturally  for the variety of 
\la\ line detections in star-forming galaxies Tenorio-Tagle et al. \cite{TKT},
have proposed a scenario based on the hydrodynamics of superbubbles powered by
 massive starbursts. This scenario is  visually depicted in
Fig.~\ref{fig:superbubble}.
The overpowering mechanical luminosity ($E_0$ $\geq$ 10$^{41}$erg s$^{-1}$)
from massive starbursts ($M_{stars}$ = 10$^5$ - 10$^6$ \Msun) is known 
to lead to a rapidly evolving superbubble able to blowout the gaseous 
central disk configuration into their extended HI haloes.
 
 As shown 
by \cite{KK} 
 the superbubble will blowout and thus massive starbursts 
will lead to  superbubble blowout phenomena even in massive galaxies 
such as the Milky Way. One can then predict  the venting of the  
hot superbubble interior gas through the 
Rayleigh-Taylor fragmented shell, into the extended HI halo
where it would  push once again the outer shock,
  quite early in the evolution of the starburst 
($T_{blowout} \leq $ 2Myr), allowing to build a new
 shell
of swept up halo matter (see Fig.~\ref{fig:superbubble}b). 


The blowout or the fragmentation of the shell  allows
the expansion of the hot interior gas into the extended low density HI halo, 
and the leakage of the $uv$ photons emitted by the starburst.
These photons establish an ionized conical 
HII region with its apex at the starburst.
 The density of the halo
steadily decreases with radius,
 and one 
can show that the conical sector of the HII region will extend all the way to 
the galaxy outer edge, i.e.~several \kpc. 
 Furthermore, 
the low halo density implies a
long recombination time  (t$_{rec}$ = 1/($\beta n_{halo}$) 
$\geq$ 10$^{7}$ yr), 
with the implication that
the ionized conical sector of the galaxy halo becomes, and remains,  
transparent to the 
ionizing flux produced  during the ensuing lifetime of the 
ionizing massive stars ($\leq$ 10 Myr).
 Then clearly, the \la\ photons
produced at the central HII region will also be able to travel freely 
in such directions. The escape of $uv$ photons from the 
galaxy would be particularly important during the early stages after
blowout
and until the new shell of swept up halo matter condenses enough 
material as to allow its recombination.
If the shock progresses with speeds of a 
few hundred ( $\leq$ 400)
km s$^{-1}$, (i.e. with a Mach number $M$ $\leq$ 40)
 it will promote the rapid cooling 
of the shocked gas
 that  will cool down to the HII region temperature 
making the shock isothermal, and thus causing  compression
factors of several hundreds.
Compression leads  to recombination in time-scales 
($t_{rec} =  1/\beta n_{shock})$ of less than 
10$^5$yr,  and this immediately
and steadily will reduce the number of stellar $uv$ photons leaking out of the 
galaxy. At the same time recombination in the fast expanding shell 
will lead to a correspondingly blueshifted \la\ emission, as depicted 
in Fig.~\ref{fig:superbubble}c.
Once the shell presents a large column density 
($\sim$ 10$^{19}$ atoms cm$^{-2}$), as it grows to dimensions of a few kpc, 
it will trap the  
ionization front.  Note that 
from then onwards,  
recombinations in the shell  will inhibit the 
further escape of ionizing photons from the galaxy (compare
 Fig.~\ref{fig:superbubble}b,
 c, 
and d).\\
The trapping of the ionization front,  makes the shell 
acquire a multiple structure with a photoionized inner edge, 
a steadily growing central zone of HI, and an outer ionized sector where 
the recently shocked 
ionized halo gas is steadily incorporated. 
The growth of the central layer eventually will cause  
sufficient scattering and absorption of the \la\ photons 
emitted by the central HII region, leading to a blueshifted \la\ 
absorption. 
As long as recombinations continue to occur at the 
leading edge of the shell, a blueshifted \la\ in emission will appear 
superposed to the 
blueshifted absorption feature (see Fig.~\ref{fig:superbubble}d).
 Recombination at the 
leading edge will become steadily less frequent, depleting the 
blueshifted \la\ in emission. This is due when the shell and its 
leading shock move into 
the outer less dense regions of the halo, and the shell recombination time, 
despite the compression at the shock, becomes larger than the dynamical time. 
At this stage, an observer looking along the conical sector of the HII region
will detect a P-Cygni-like \la\ line profile as shown in
 Fig.~\ref{fig:superbubble}e. \\

Geometrical dilution 
of the $uv$ flux will begin to make an impact as the superbubble grows 
large. This 
and the drop in the $uv$ photon production rate, 
caused by the death of the most massive stars after  
$t$ = $t_{ms}$, will enhance the
column density of neutral material in the central zone of the 
recombined shell to eventually cause the full saturated 
absorption of the \la\ line (see Fig.~\ref{fig:superbubble}f).
 Full saturated absorption has usually been accounted for by the large
column 
density of the extended HI envelope of these galaxies and thus, as in all 
models, many different orientations will match the observations.


\section{Discussion} 

\subsection{The relevance of the superbubble model}
The main implication of the evolution depicted in Fig.~\ref{fig:superbubble}
is that it is the feedback from the 
massive stars that 
leads to  the large variety of \la\ emission profiles. 
P-Cygni \la\ profiles are predicted when 
observing along the angle subtended by the conical HII region but only once 
the ionization front is trapped by the sector of the superbubble shell
evolving into the extended halo. This will produce the fast moving 
layer of HI at the 
superbubble shell, here thought to be responsible for 
the partial absorption observed in sources such as Haro 2, ESO 400-G043 (which
probably exhibits a secondary blueshifted \la\ emission) and
 ESO 350-IG038 \cite{K98}.

Damped \la\ absorption is seen in several galaxies. We note that 
these objects are all gas--rich dwarf galaxies whereas in most cases but 
Haro 2, the  HII galaxies that exhibit \la\ in emission or
 with a P-Cygni profile,
are on the higher luminosity side of the distribution ($M$ $\leq$ $-18$).

Pure \la\ emission is observed in C0840+1201 and T1247-232 
(\cite{TDT}; IUE)  or T1214-277 (\cite{TI}; HST). 
Such a line implies no absorption and thus no HI gas between the starburst 
HII region and the observer, as when observing the 
central HII region after the superbubble blowout, within the conical HII
region carved in the extended HI halo. 
It is not a straigthforward issue to estimate what is the 
fraction of Lyman continuum 
radiation that leaks out from galaxies.
 This scenario 
however predicts a short but significant evolutionary 
phase (between blowout and the trapping of the ionization front 
by the fast expanding shell) during which a large amount of  $uv$ 
radiation 
could leak out of a galaxy into the intergalactic medium. 
Detailed numerical calculations of the scenario proposed here are
currently underway. These results and further implications of the model 
will be reported in a forthcoming communication.

\subsection{The Galaxies at High-Redshift}

The effect of neutral gas flows helps to understand why luminous
high-redshift objects have only been found up to now with linewidths
larger than
1000 km s$^{-1}$.  High--redshift galaxies with very strong (EWs $>$
500~\AA) extended Ly$\alpha$\ emission are characterized by strong velocity
 shears
and turbulence (v $>$ 1000 km s$^{-1}$); this suggests an AGN activity,
in the sense that other ISM energising mechanism than photoionization by
young stars may be operating. On the other hand Steidel et al. \cite{SGP}
 have recently
discovered a substantial population of star--forming galaxies at
3.0$<$z$<$3.5 that were selected 
but from the presence of a very blue far-UV continuum and a break below
912~\AA\ in the rest frame. These Lyman--break galaxies (LBG) are much more
luminous  but similar
 to our local starbursts in the sense that
50\%  show no Ly$\alpha$\ emission whereas the rest does, but
 with weak EWs no larger than 20~\AA\ at rest.  The Ly$\alpha$\ profiles of
 this
population looks also very similar to those of our local starburst
galaxies (\cite{F,PS}; see also \cite{DSS} for the z=5.34 galaxy).\\ 
New  \la\ emitters are now found at high-redshift from surveys using
large telescopes with narrow-band filters (\cite{HC,PW}). Limits down
to  a few 10$^{-18}$ erg/cm$^{2}$/sec are now reachable and give access, in
principle, to galaxies with fainter continuum magnitudes than the LBG. It is not
clear whether these Lyman--alpha--galaxies (LAG)  represent the earliest
stages of the galaxy formation process as suggested by Hu et al. \cite{HC}.
Are these galaxies less dusty than the LBGs? Is there an intrinsically fainter
population of dust-free star--forming galaxies,
 possibly connected to the bright-end of the star-forming dwarfs
population (-18 to -20?) we see today? What are their masses and spatial
distribution (see also \cite{PW} for a recent HST study)? 
 Near-IR spectroscopic studies with large ground based telescopes
 will in the future help  to settle some of these questions \cite{PPP}.\\
The  possible increase of the star formation rates
as one moves at higher redshift (between 3 and 6) reported by
 Hu et al. \cite{HC} must be taken with 
caution since from the preceeding
 discussion, it is clear that a significant, yet unknown fraction of
 the youngest galaxies may not be \la\ emitters. Indeed, since  \la\
is primarely a diagnostic of the ISM, it is only when we fully
understand how the ISM and \la\ are related can we hope to firmly relate
\la\ to the SFR.\\

\vfill

\end{document}